\documentclass[11pt]{article}
\usepackage{amssymb}\usepackage{color}\usepackage[colorlinks]{hyperref}

\usepackage{tikz}
\usepackage{stmaryrd}\usepackage{mathrsfs}
\usepackage{graphicx}
\usepackage{amsmath}
\usepackage{caption}
\usepackage{subcaption}
\usepackage{listings}
\usepackage{verbatim}
\usepackage{braket}
\usepackage{subfiles}
\usepackage{bm}

\begin{document}
	\setlength{\captionmargin}{27pt}
	\newcommand\hreff[1]{\href {http://#1} {\small http://#1}}
	\newcommand\trm[1]{{\bf\em #1}} \newcommand\emm[1]{{\ensuremath{#1}}}
	\newcommand\prf{\paragraph{Proof.}}\newcommand\qed{\hfill\emm\blacksquare}
	
	\newtheorem{thr}{Theorem} 
	\newtheorem{lmm}{Lemma}
	\newtheorem{cor}{Corollary}
	\newtheorem{con}{Conjecture} 
	\newtheorem{prp}{Proposition}
	
	\newcommand\QC{\mathbf{QC}} 
	\newcommand\C{\mathbf{C}} 

	\newtheorem{blk}{Block}
	\newtheorem{dff}{Definition}
	\newtheorem{asm}{Assumption}
	\newtheorem{rmk}{Remark}
	\newtheorem{clm}{Claim}
	\newtheorem{exm}{Example}
	\newtheorem{exc}{Exercise}
	
	\newcommand\Ks{\mathbf{Ks}} 
	\newcommand{\ab}{a\!b}
	\newcommand{\yx}{y\!x}
	\newcommand{\yux}{y\!\underline{x}}
	
	\newcommand\floor[1]{{\lfloor#1\rfloor}}
	\newcommand\ceil[1]{{\lceil#1\rceil}}
	
	\renewcommand\do[1]{\overline{\overline{#1}}}

	\newcommand{\bmu}{\boldsymbol{\mu}}
	\newcommand{\bnu}{\boldsymbol{\nu}}
	\newcommand{\lea}{<^+}
	\newcommand{\gea}{>^+}
	\newcommand{\eqa}{=^+}

	\newcommand{\lel}{<^{\log}}
	\newcommand{\gel}{>^{\log}}
	\newcommand{\eql}{=^{\log}}
	
	\newcommand{\F}{\mathbf{F}}
	\newcommand{\E}{\mathbf{E}}
	\newcommand{\lem}{\stackrel{\ast}{<}}
	\newcommand{\gem}{\stackrel{\ast}{>}}
	\newcommand{\eqm}{\stackrel{\ast}{=}}
    \newcommand{\Huc}{\mathbf{Huc}}
	\newcommand\edf{{\,\stackrel{\mbox{\tiny def}}=\,}}
	\newcommand\edl{{\,\stackrel{\mbox{\tiny def}}\leq\,}}
	\newcommand\then{\Rightarrow}

	\newcommand{\kpsi}{\ket{\psi}}
	\newcommand{\ktheta}{\ket{\theta}}
	
	\newcommand\Ip{\I_\mathrm{Prob}}
	\newcommand\uhr{\upharpoonright}
	\newcommand\Hg{\mathbf{Hg}}
	\newcommand\Hv{\mathbf{Hv}}
	\newcommand\Hbvl{\mathbf{Hbvl}}
	\newcommand\Hb{\mathbf{Hb}}
	\renewcommand\H{\mathbf{H}}
	\newcommand\ml{\underline{\mathbf m}}
	\newcommand\mup{\overline{\mathbf m}}
	\newcommand\UI{\mathcal{I}}
	\newcommand\km{{\mathbf {km}}}\renewcommand\t{{\mathbf {t}}}
	\newcommand\KM{{\mathbf {KM}}}\newcommand\m{{\mathbf {m}}}
	\newcommand\md{{\mathbf {m}_{\mathbf{d}}}}\newcommand\mT{{\mathbf {m}_{\mathbf{T}}}}
	\newcommand\K{{\mathbf K}} \newcommand\I{{\mathbf I}}
	
	\newcommand\II{\hat{\mathbf I}}
	\newcommand\Kd{{\mathbf{Kd}}} \newcommand\KT{{\mathbf{KT}}} 
	\renewcommand\d{{\mathbf d}} 
	\newcommand\D{{\mathbf D}}
	\newcommand\Tr{\mathrm{Tr}}
	\newcommand\w{{\mathbf w}}
	\newcommand\Cs{\mathbf{Cs}} \newcommand\q{{\mathbf q}}
	\newcommand\St{{\mathbf S}}
	\newcommand\M{{\mathbf M}}\newcommand\Q{{\mathbf Q}}
	\newcommand\ch{{\mathcal H}} \renewcommand\l{\tau}
	\newcommand\tb{{\mathbf t}} \renewcommand\L{{\mathbf L}}
	\newcommand\bb{{\mathbf {bb}}}\newcommand\Km{{\mathbf {Km}}}
	\renewcommand\q{{\mathbf q}}\newcommand\J{{\mathbf J}}
	\newcommand\z{\mathbf{z}}
	\newcommand\Z{\mathbb{Z}}
	\newcommand\Hn{\mathbf{Hn}}

	\newcommand\B{\mathbf{bb}}\newcommand\f{\mathbf{f}}
	\newcommand\hd{\mathbf{0'}} \newcommand\T{{\mathbf T}}
	\newcommand\R{\mathbb{R}}\renewcommand\Q{\mathbb{Q}}
	\newcommand\N{\mathbb{N}}\newcommand\BT{\{0,1\}}
	\newcommand\W{\mathbb{W}}
	\newcommand\dom{\mathrm{Dom}}
	\newcommand\FS{\BT^*}\newcommand\IS{\BT^\infty}
	\newcommand\FIS{\BT^{*\infty}}
	\renewcommand\S{\mathcal{C}}\newcommand\ST{\mathcal{S}}
	\newcommand\UM{\nu_0}\newcommand\EN{\mathcal{W}}
	
	\newcommand{\supp}{\mathrm{Supp}}
	
	\newcommand\lenum{\lbrack\!\lbrack}
	\newcommand\renum{\rbrack\!\rbrack}
	
	\newcommand\om{\overline{\mu}}
	\newcommand\on{\overline{\nu}}
	\newcommand\h{\mathbf{h}}
	\renewcommand\qed{\hfill\emm\square}
	\renewcommand\i{\mathbf{i}}
	\newcommand\p{\mathbf{p}}
	\renewcommand\q{\mathbf{q}}
	\renewcommand\T{\mathbf{T}}
	
	\title{Some Implications of the Independence Postulate for Physics}
	
	\author {Samuel Epstein\\samepst@jptheorygroup.org}
	\maketitle
  \begin{abstract}
 The Many Worlds Theory and Constructor Theory are in conflict with the Independence Postulate. The conflict with the Many Worlds Theory is shown through the existence of a finite experiment that measures the spin of a large number of electrons.  After the experiment there are branches of positive probability which contain forbidden sequences that break the Independence Postulate. Constructor Theory consists of counterfactuals, decreeing certain processes can or cannot occur. However this binary classification meets challenges when describing whether a forbidden sequence can be found or created.
 \end{abstract}

%      
% \end{abstract}
 
The Many Worlds Theory (\textbf{MWT}) was formulated by Hugh Everett \cite{Everett57} as a solution to the measurement problem of Quantum Mechanics. Branching (i.e. splitting of worlds) occurs during any process that magnifies microscopic superpositions to the macro-scale. This occurs in events including human measurements such as the double slit experiments, or natural processes such as radiation resulting in cell mutations. 

One question is if \textbf{MWT} causes issues with the foundations of computer science. The physical Church Turing Thesis (\textbf{PCTT}) states that any functions computed by a physical system can be simulated by a Turing machine.  A straw man argument for showing \textbf{MWT} and \textbf{PCTT} are in conflict is an experiment that measures the spin of an unending number of electrons, with each measurement bifurcating the current branch into two sub-branches. This results in a single branch in which the halting sequence is outputted. However this branch has Born probability converging to 0, and can be seen as a deviant, atypical branch.

In fact, conflicts do emerge between \textbf{MWT} and  Algorithmic Information Theory. In particular, the Independence Postulate (\textbf{IP}) \cite{Levin84,Levin13} is a finitary Church-Turing thesis, postulating that certain infinite and \textit{finite} sequences cannot be found in nature, a.k.a. have high ``addresses''. One such set of forbidden sequences are large prefixes of the halting sequence. If a forbidden sequence is found in nature, an information leak will occur. However \textbf{MWT} represents a theory in which such information leaks can occur. 

Another promising area of research is Constructor Theory (\textbf{CT}) \cite{Deutsch13} with main proponents David Deutsch and Chiara Marletto. \textbf{CT} aims to unify many areas of science with counterfactuals. Counterfactuals describe which processes that can occur or not occur. These counterfactuals are principles which it is conjectured that all laws of physics must adhere to. The basis tenet of \textbf{CT} is \cite{DeutschMa15}
\begin{quote}
\textit{All other laws of physics are expressible entirely in terms of statements about which physical transformations are possible and which are impossible, and why.}
\end{quote}

In an online colloquium \cite{Deutsch2022}, David Deutsch was asked if G\"{o}del's Incompleteness Theorem or the halting problem would be incorporated into \textbf{CT}. David Deutsch responded with:
\begin{quote}
    \textit{``No they wouldn't, at least we don't expect them to be added because those issues only arise in infinite sets and constructor theory regards\dots physical systems as always finite. It only makes statements about finite systems.''}
\end{quote}
However, among other things, $\textbf{IP}$ is a finitary version of the halting problem, so we believe it must be reconciled with \textbf{CT}. The main issue is the following question: Is it possible or impossible to create or find a large prefix of the halting sequence? Is this even a well formed (answerable) question? For example, it is possible to create one such large prefix $h$ if one were to find any $y$ of the same length and $y\oplus h$. But obviously such a construction seems lacking.

\section{The Independence Postulate}
\label{sec:ip}
In \cite{Levin84,Levin13}, the Independence Postulate, \textbf{IP}, was introduced:
\begin{quote}
	\textit{Let $\alpha\in\FIS$ be a sequence defined with an $n$-bit mathematical statement (e.g., in Peano Arithmetic or Set Theory), and a sequence $\beta\in\FIS$ can be located in the physical world with a $k$-bit instruction set (e.g., ip-address). Then $\I(\alpha : \beta) < k+n+c$, for some small absolute constant $c$.}
\end{quote}
The $\I$ term is an information measure in Algorithmic Information Theory. For this chapter, the information term used is $\I(x:y)=\K(x)+\K(y)-\K(x,y)$, where $\K$ is the prefix-free Kolmogorov complexity. This definition of $\I$ can be used  because the thought experiment only deal with finite sequences. 

Let $\Omega_m$ be the first $m$ bits of Chaitin's Omega (the probability that a universal Turing machine will halt). It is well known that $m\lea \K(\Omega_m)$. Furthermore $\Omega_m$ can be described by a mathematical formula of size $O(\log m)$. Thus by \textbf{IP}, where $\Omega_m=\alpha=\beta$, $\Omega_m$ can only be found with physical addresses of size at least $m-O(\log m)$. Thus finding any sufficiently large sequence $\Omega_m$ is not physically possible. This is due to fact that the observable universe is $8.8\times 10^{26}$ meters across and a transistor can only be made to be $2\times 10^{-9}$ meters long. Thus if the minimum length of an address for a sequence is greater than a thousand, it cannot exist in nature. This sentiment was reflected in \cite{Levin13}, where sequences with small addresses are called ``physical'', and thus sequences with only high addresses are ``unphysical''. As we shall see in the next parts of this paper, the sequence $\Omega_m$ for large enough $m$ will cause trouble for both \textbf{MWT} and \textbf{CT}.

\section{Many Worlds Theory}

Some researchers believe there is an inherent problem in quantum mechanics. On one hand, the dynamics of quantum states is prescribed by unitary evolution. This evolution is deterministic and linear. On the other hand, measurements result in the collapse of the wavefunction. This evolution is non-linear and nondeterministic. This conflict is called the measurement problem of quantum mechanics.

The time of the collapse is undefined and the criteria for the kind of collapse are strange. The Born rule assigns probabilities to macroscopic outcomes. The projection postulate assigns new microscopic states to the system measured, depending on the the macroscopic outcome.
One could argue that the apparatus itself should be modeled in quantum mechanics. However it's dynamics is deterministic. Probabilities only enter the conventional theory with the measurement postulates.

\textbf{MWT}  was proposed by Everett as a way to remove the measurement postulate from quantum mechanics. The theory consists of unitary evolutions of quantum states without measurement collapses. For \textbf{MWT}, the collapse of the wave function is the change in dynamical influence of one part of the wavefunction over another, the decoherence of one part from the other. The result is a branching structure of the wavefunction and a collapse only in the phenomenological sense.

\subsection{Branching Worlds}
An example of a branching of universes can be seen in an idealized experiment with a single electron with spin $\ket{\phi_\uparrow}$ and $\ket{\phi_\downarrow}$. This description can be found in \cite{saundersBaKeWa2010}. There is a measuring apparatus $\mathcal{A}$, which is in an initial state of $\ket{\psi^\mathcal{A}_\mathrm{ready}}$. After $\mathcal{A}$ reads spin-up or spin-down then it is in state $\ket{\psi^\mathcal{A}_{\textrm{reads spin }\uparrow}}$ or $\ket{\psi^\mathcal{A}_{\textrm{reads spin }\downarrow}}$, respectively. The evolution for when the electron is solely spin-up or spin-down is
\begin{align*}
\ket{\phi_\uparrow}\otimes \ket{\psi^\mathcal{A}_\mathrm{ready}} &\stackrel{\mathrm{unitary}}{\longrightarrow} \ket{\phi_\uparrow}\otimes \ket{\psi^\mathcal{A}_{\textrm{reads spin }\uparrow}}\\
\ket{\phi_\downarrow}\otimes \ket{\psi^\mathcal{A}_\mathrm{ready}} &\stackrel{\mathrm{unitary}}{\longrightarrow} \ket{\phi_\downarrow}\otimes \ket{\psi^\mathcal{A}_{\textrm{reads spin }\downarrow}}.
\end{align*}
Furthermore, one can model the entire quantum state of an observer $\mathcal{O}$ of the apparatus, with

\begin{align*}
&\ket{\phi_\uparrow}\otimes \ket{\psi^\mathcal{A}_\mathrm{ready}}\otimes \ket{\xi^\mathcal{O}_\mathrm{ready}} \\
\stackrel{\mathrm{unitary}}{\longrightarrow}& \ket{\phi_\uparrow}\otimes \ket{\psi^\mathcal{A}_{\textrm{reads spin }\uparrow}}\otimes \ket{\xi^\mathcal{O}_\mathrm{ready}}\\
\stackrel{\mathrm{unitary}}{\longrightarrow}& \ket{\phi_\uparrow}\otimes \ket{\psi^\mathcal{A}_{\textrm{reads spin }\uparrow}}\otimes \ket{\xi^\mathcal{O}_{\textrm{reads spin }\uparrow}}\\
\\
&\ket{\phi_\downarrow}\otimes \ket{\psi^\mathcal{A}_\mathrm{ready}}\otimes \ket{\xi^\mathcal{O}_\mathrm{ready}}\\ \stackrel{\mathrm{unitary}}{\longrightarrow}& \ket{\phi_\downarrow}\otimes \ket{\psi^\mathcal{A}_{\textrm{reads spin }\downarrow}}\otimes \ket{\xi^\mathcal{O}_\mathrm{ready}}\\
\stackrel{\mathrm{unitary}}{\longrightarrow}& \ket{\phi_\downarrow}\otimes \ket{\psi^\mathcal{A}_{\textrm{reads spin }\downarrow}}\otimes \ket{\xi^\mathcal{O}_{\textrm{reads spin }\downarrow}}.\\
\end{align*}
For the general case, the electron is in a state $\ket{\phi} = a\ket{\phi_\uparrow}+b\ket{\phi_\downarrow}$, where $|a|^2+|b|^2=1$. In this case, the final superposition would be of the form:
\begin{align*}
&a\ket{\phi_\uparrow}\otimes \ket{\psi^\mathcal{A}_{\textrm{reads spin }\uparrow}}\otimes \ket{\xi^\mathcal{O}_{\textrm{reads spin }\uparrow}}\\
+&b\ket{\phi_\downarrow}\otimes \ket{\psi^\mathcal{A}_{\textrm{reads spin }\downarrow}}\otimes \ket{\xi^\mathcal{O}_{\textrm{reads spin }\downarrow}}.
\end{align*}
This is a superposition of two branches, each of which describes a perfectly reasonable physical story. This bifurcation is one method on how the  quantum state of universe bifurcates into two branches.
\subsection{Deriving the Born Rule}
\label{sec:prob}
In the author's opinion, one of the main problems of \textbf{MWT} is its reconciliation of the Born rule, for which no proposed solution has universal consensus. In standard quantum mechanics, measurements are probabilistic operations. Measurements on a state vector $\ket{\psi}$, which is a unit vector over Hilbert space $\mathcal{H}$, are self-adjoint operators $\mathcal{O}$ on $\mathcal{H}$. Observables are real numbers that are the spectrum $\mathrm{Sp}(\mathcal{O})$ of $\mathcal{O}$. A measurement outcome is a subset $E\subseteq\mathrm{Sp}(\mathcal{O})$ with associated projector $P_E$ on $\mathcal{H}$. Outcome $E$ is observed on measurement of $\mathcal{O}$ on $\ket{\psi}$ with probability $P(E)=\bra{\psi}P_E\ket{\psi}$. This is known as the Born rule. After this measurement, the new state becomes $P_E\ket{\psi}{/}\sqrt{\bra{\psi}P_E\ket{\psi}}$. This is known as the projection postulate.

However, the Born rule and the projection postulate are not assumed by \textbf{MWT}. The dynamics are totally deterministic. Each branch is equally real to the observers in it. To address these issues, Everett first derived a typicality-measure that weights each branch of a state's superposition. Assuming a set of desirable constraints, Everett derived the typicality-measure to be equal to the norm-squared of the coefficients of each branch, i.e. the Born probability of each branch. Everett then drew a distinction between typical branches that have high typicality-measure and exotic atypical branches of decreasing typicality-measure. For the repeated measurements of the spin of an electron $\ket{\phi}=a\ket{\phi_\uparrow}+b\ket{\phi_\downarrow}$, the relative frequencies of up and down spin measurements in a typical branch converge to $|a|^2$ and $|b|^2$, respectively. The notion of typicality can be extended to measurements with many observables.

In a more recent resolution to the relation between \textbf{MWT} and probability, Deutsch introduced a decision theoretic interpretation \cite{Deutsch99} that obtains the Born rule from the non-probabilistic axioms of quantum theory and non-probabilistic axioms of decision theory. Deutsch proved that rational actors are compelled to adopt the Born rule as the probability measure associated with their available actions. This approach is highly controversial, as some critics say the idea has circular logic.

Another attempt uses subjective probability \cite{Vaidman98}. The experimenter puts on a blindfold before he finishes performing the experiment. After he finishes the experiment, he has uncertainty about what world he is in. This uncertainty is the foundation of a probability measure over the measurements. However, the actual form of the probability measure needs to be postulated:\\

\noindent\textbf{Probability Postulate.} \textit{An observer should set his subjective probability of the outcome of a quantum experiment in proportion to the total measure of existence of all worlds with that outcome.}\\

Whichever explanation of the Born rule one adopts, the following section shows there is an issue with \textbf{MWT} and \textbf{IP}. There exist branches of substantial Born probability where information leaks occurs. 

\section{Violating the Independence Postulate}
\label{sec:ip}
\textbf{IP} can be violated in the following idealized  experiment measuring the spin $\ket{\phi_\uparrow}$ and $\ket{\phi_\downarrow}$ of $m$ isolated electrons. We denote $\ket{\phi_0}$ for $\ket{\phi_\uparrow}$ and $\ket{\phi_1}$ for $\ket{\phi_\downarrow}$. The “address” (in the sense of \textbf{IP}) of this experiment (such as the physical address of the Large Hadron Collider) is $< O(\log m)$.  The measuring apparatus will measure the spin of $m$ electrons in the state $\ket{\phi}=\frac{1}{2}\ket{\phi_\uparrow}+\frac{1}{2}\ket{\phi_\downarrow}$. There is a measuring apparatus $\mathcal{A}$ with initial state of $\ket{\psi^\mathcal{A}}$, and after reading $m$ spins of $m$ electrons, it is in the state $\ket{\psi^\mathcal{A}[x]}$, where $x\in\BT^m$, whose $i$th bit is 1 iff the $i$th measurement returns $\ket{\phi_1}$. 

 The experiment evolves according to the following unitary transformation:

\begin{align*}
&\bigotimes_{i=1}^m\ket{\phi}\otimes \ket{\psi^\mathcal{A}} 
\stackrel{\mathrm{unitary}}{\longrightarrow}\sum_{a_1,\dots,a_m\in\BT^m}2^{-m/2}\bigotimes_{i=1}^m\ket{\phi_{a_i}}\otimes\ket{\psi^\mathcal{A}[a_1a_2\dots a_m]}.\\
\end{align*}

If the bits returned are $\Omega_m$ then a memory leak of size $m-O(\log m)$ has occurred, because $\Omega_m$ has been located by the address of the experiment, which is $O(\log m)$. Thus
$$\textrm{Born-Probability}(\textrm{a memory leak of size $m-O(\log m)$ occurred}) \geq 2^{-m}.$$

\section{Reconciling \textbf{MWT} and \textbf{IP}}
\label{sec:reconmwtip}
There are multiple variations of \textbf{MWT} when it comes to consistency across universes. In one formulation, all universes conform to the same physical laws. In another model, each universe has its own laws, for example different values of gravitational constant, etc. However, the experiment in the previous section shows that mathematics itself is different between universes, regardless of which model is used. In some universes, \textbf{IP} holds and there is no way to create information leaks. In other universes information leaks occur, and there are tasks where randomized algorithms fail but non-algorithmic physical methods succeeds. One such task is finding new axioms of mathematics. This was envisioned as a possibility by G\"{o}del \cite{Godel61}, but there is a universal consensus of the impossibility of this task. Not any more! In addition, because information leaks are finite events, the Born probability of worlds containing them is not insignificant. In such worlds, \textbf{IP} cannot be formulated, and the the foundations of Algorithmic Information Theory itself become detached from reality. 

Formulated another way, let us suppose the Born probability is derived from the probability postulate. We have a ``blindfolded mathematician'' who performs the experiment described above. Before the mathematician takes off her blindfold, she states the Independence Postulate. By the probability postulate, with measure $2^{-m}$ over all worlds, there is a memory leak of size $m-O(\log m)$ and \textbf{IP} statement by the mathematician is in error.

\subsection{The Probability Rebuttal}
As a rebuttal, one can, with non-zero probability, just flip a coin $N$ times and get $N$ bits of Chaitin’s Omega. Or more generally, how does one account for a probability $P$ over finite or infinite sequences learning information about a forbidden sequence $\beta$ with good probability? Due to probabilistic conservation laws \cite{Levin74,Levin84}, we have
$$
\Pr_{\alpha\sim P}[\I(\alpha:\beta)>\I(\langle P\rangle:\beta)+m]\lem 2^{-m}.
$$
Thus the probability of a single event creating a leak is very small. However if many events occur, then the chances of a memory leak grows. However as there is many events, to locate one such leak, one will probably need a long address to find the leak, balancing out the \textbf{IP} equation.

This still leaves open the possibility of a memory leak occurring at an event with a small address. For example, say someone assigns a random 1 trillion bit sequence to every ip-address. What are the chances that one of them is $\Omega_{10^{12}}$? Since there are a small number of events that have a small address, the probability of a significant memory leak is extremely small. In physics on can postulate away events with extremely small probabilities. For example, the second law of thermodynamics states that entropy is non-decreasing, postulating away the extremely unlikely event that a large system suddenly decreases in thermodynamic entropy, i.e. a broken vase forming back to together.

\subsection{Memory Leaks}
There is no way to postulate away such memory leaks in \textbf{MWT}. Assuming the \textit{probability postulate}, probability is a measure over the space of possible worlds. Thus when Bob now threatens to measure the spin of $m$ particles, Alice now knows $2^{-m}$ of the resultant worlds will contain $m$ bits of Chaitin’s Omega, violating \textbf{IP}.

\section{Constructor Theory}

\textbf{CT} aims to define a set of principles, or counterfactuals that constrain how laws of the physics. For more motivation on \textbf{CT}, I refer the readers to \cite{Deutsch13, Marletto21}. Fundamental in \textbf{CT} are constructors which change substrates,
$$
\textrm{Input Substrate} \xrightarrow{\textrm{Constructor}}\textrm{Output Substrate.}
$$
A \textit{constructor task} $\mathfrak{A}$ is a set of pairs each designating an input state for the task and an output state for that input. For example,
$$
\mathfrak{A}=\{x_1\rightarrow y_1, x_2\rightarrow y_2,\dots\}.
$$

A constructor is capable of performing task $\mathfrak{A}$, if whenever it is given a substrate in a input attribute of $\mathfrak{A}$, it transforms them to one of the output attribute that $\mathfrak{A}$ associates with that input. 
\begin{dff}
\label{dff:taskimp}
A task $\mathfrak{A}$ is \textit{impossible} if there is a law of physics that forbids it being carried out with arbitrary accuracy and reliably by a constructor. Otherwise,  $\mathfrak{A}$ is \textit{possible}, with $\mathfrak{A}^\checkmark$. 
\end{dff}
However a possible task is not guaranteed to occur. The central tenet of \textbf{CT} is that the laws of physics must conform to a set of counterfactuals or principles which are statements on  whether tasks are possible or impossible. \textbf{CT} has been applied to large number of areas, including (but not limited to) classical information, quantum information theory, probability, life, and thermodynamics \cite{Deutsch13,DeutschMa15,Marletto15,Marletto16}. In the next section we will review the intersection of \textbf{CT} and information.

\subsection{Information}
In my opinion, the most successful endeavour of \textbf{CT} is its application to classical and quantum information. We review the work in \cite{DeutschMa15}. Any set of disjoint attributes is called a \textit{varriable}. Whenever a substrate is in a state with attribute $x\in X$, where $X$ is a variable, we say $X$ is sharp with value $x$. Einstein's (1949) principle of locality has an intepretation as a counterfactual \cite{DeutschMa15},
\begin{quote}
    \textit{There exists a mode of description such that the state of the combined system $\mathbf{S}_1\oplus \mathbf{S}_2$ of any two substrates $\mathbf{S}_1$ and $\mathbf{S}_2$ is the ordered pair $(x,y)$ of the states $x$ of $\mathbf{S_1}$ and $y$ of $\mathbf{S}_2$, and any construction undergone by $\mathbf{S}_1$ and not $\mathbf{S}_2$ can change only $x$ and not $y$.}
\end{quote}
The theory of information relies on a \textbf{CT} theoretic description of \textit{computation}. A \textit{reversible computation} $\mathfrak{C}_\Pi(S)$ is the task of performing a permutation $\Pi$ over some set $S$ of at least two possible attributes of some substrate:
$$
\mathfrak{C}_\Pi(S)=\bigcup\left\{x\rightarrow \Pi(x)\right\}.
$$
A \textit{computation variable} is a set $S$ of two or more possible attrbutes for which $\mathfrak{C}^\checkmark_\Pi$ for all permutation $\Pi$ over $S$. A \textit{computation medium} is a substrate with at least one computation variable. With computation formalized, we are now ready to define information. 

The \textit{cloning task} for a set $S$ of possible attributes of substrate $\mathbf{S}$ is the task
$$
\mathfrak{R}_s(x_0)=\bigcup\{(x,x_0)\rightarrow (x,x)\}.
$$
on substrate $\mathbf{S}\oplus\mathbf{S}$, where $x_0$ is some attribute realizable from naturally occurring resources. We will revisit this statement later. An \textit{information variable} is a cloneable computation variable. An \textit{information attribute} is a member of an information variable, and an \textit{information medium} is a substrate that has at least one information variable. 

A set $X$ of possible attributes of a substrate \textbf{S} is \textit{distinguishable} if
$$
\left(\bigcup_{x\in X}\left\{x\rightarrow \Psi_x\right\}\right)^\checkmark,
$$
where $\left\{\Psi_x\right\}$ is a information variable. If the original substrate continues to exist and the process stores its result in a second \textit{output substrate}, the input variable $X$ is \textit{measurable}:
$$
\left(\bigcup_{x\in X}\left\{(x,x_0)\rightarrow (y_x,`x'\right\}\right)^\checkmark.
$$
The output substrate is prepared with a `receptive' attribute $x_0$. We introduce the task of preparing a substrate. A variable $X$ in a substrate \textbf{S} is \textit{preparable} if there is a information medium $\textbf{R}$ with an information variable $W$ and a possible task $\mathfrak{A}^\checkmark$ such that for all $x\in X$ there is a $w(x)\in W$ such that $\{w(x)\rightarrow x\}\in \mathfrak{A}$. Thus
$$
\left(\bigcup_{x\in X}\left\{w(x)\rightarrow x\right\}\right)^\checkmark.
$$
\section{Reconciling \textbf{CT} and \textbf{IP}}
We discuss three issues between \textbf{IP} and \textbf{CT}. First, we recall from Section \ref{sec:ip}, that $\Omega_m$ is a forbidden string, and can only be found with a physical address of size at least $m-O(\log m)$. However the following principle \cite{DeutschMa15} states
\begin{quote}
\textit{\textbf{VI}. Any number of instance of any information medium, with any one of its information-instantiating attributes, is preparable from naturally occurring substrates.}
\end{quote}
Take the information medium of a computer that has information variable $\BT^m$, for very large (but finite) $m$. By Principle \textbf{VI}, $\Omega_m$ can be prepared by naturally occurring substrates. However $\textbf{IP}$ postulates that all such $\Omega_m$ cannot be found without an address almost the size of $\Omega_m$. There is a difference in language from ``naturally occurring substrates'' and ``high address'' but the conflict remains. 

We point out the second issue. An information variable $S$ is a (cloneable) computation variable, such that every task permuting $S$ with permutation $\Pi$ is possible. Let the information variable $S$ consist of all strings of length $m$ for very large $m$. Let $\Pi$ be a permutation that sends $0^m$ to $\Omega_m$. By definition \ref{dff:taskimp}, this task is possible because there are no laws  of physics which forbid this transformation to occur. However if this task is possible, $0^m$ is an address that can be compressed to a $O(\log m)$ address for $\Omega_m$, contradicting $\textbf{IP}$. Thus by \textbf{IP}, no information variables can exist.

We now point out the third issue. Let us say there is a task $\mathfrak{B}$ that maps a \textit{preparable} variable $S$ in a substrate \textbf{S} to a \textit{measurable} variable $T$ in substrate \textbf{T}. Since $S$ is preparable, there is an information medium \textbf{R} with information variable $R$ that can be mapped onto $X$ with possible task $\mathfrak{A}$. Furthermore, since $T$ is measurable, it can be mapped onto an information variable $U$ in an information medium \textbf{U} with possible task $\mathfrak{C}$. In \textbf{CT}, one must conclude either the task $\mathfrak{A}$ is possible or impossible. Assume $\mathfrak{A}$ is possible. Then the combined task
$$
(\mathfrak{ABC})^\checkmark
$$
is possible. However if $\mathfrak{ABC}$ sends $0^m$ to $\Omega_m$, then by \textbf{IP}, the combined task is impossible, and thus task $\mathfrak{B}$ is impossible. So now the question of possibility or impossibility of a task must take into account whether it can be combined with a preparer and a measurer to create an information leak. This is an intractable question that applies to every task with preparable input variables and measurable output variables.
 
\subsection{The Halting Sequence Revisited}
We revisit the question posed in the introduction:
\begin{quote}
    \textit{Is it possible or impossible to create or find a large prefix of the halting sequence?}
\end{quote}
If we assert that this task is \textit{possible}, then a violation of \textbf{IP} occurs. If we assert that this task is \textit{impossible}, then this statement will be in conflict with how information variables are defined in \textbf{CT}. Thus there is a definite conflict between \textbf{IP} and \textbf{CT}. 

\section{Conclusion}

As discussed in Section \ref{sec:reconmwtip}, \textbf{IP} postulates away a union of ``bad'' events. Such ``forbidden'' events break the inequality of \textbf{IP} and were initially called ``information leaks''. One can postulate away such leaks because the probability of single leak occurring is astronomically small \cite{Levin13}. It remains to be seen how to reconcile \textbf{MWT}, \textbf{CT}, and \textbf{IP}. The simplest way to reconcile the \textbf{MWT} and \textbf{IP} is to just acknowledge there are branches where \textbf{IP} fails. Similarly, the easiest way to reconcile \textbf{CT} and \textbf{IP} is to discard one of them. Otherwise one would need statements like $\mathfrak{A}^{\checkmark*}$, which means that task $\mathfrak{A}$  is possible, but where $*$ is some additional theoretical equipment, such as an address system or a condition that a memory leak does not occur. However this reconciliation would be cumbersome, spoiling the elegance of \textbf{CT}. It remains to be seen how to overcome these obstacles.

%\bibliographystyle{alpha} 
%\bibliography{refnotes}
 
\end{document}